# Quality of Experience (QoE) beyond Quality of Service (QoS) as its baseline: QoE at the Interface of Experience Domains

**Reza FARRAHI MOGHADDAM** · **Mohamed CHERIET**

**Abstract** In this work, a new approach to the definition of the quality of experience is presented. By considering the quality of service as a baseline, that portion of the QoE that can be inferred from the QoS is excluded, and then the rest of the QoE is approached with the notion of QoE at a Boundary (QoEaaB). With the QoEaaB as the core of the proposed approach, various potential boundaries, and their associated unseen opportunities to improve the QoE are discussed. In particular, property, contract, SLA, and content are explored in terms of their boundaries and also their associated QoEaaB. With an interest in online video delivery, management of resource sharing and isolation associated with multi-tenant operations is considered. It is concluded that the proposed QoEaaB can bring a new perspective in QoE modeling and assessment toward a more enriched approach to improving the experience based on innovation and deep connectivity among actors.

**Keywords** Quality of Experience (QoE) · Quality of Service (QoS) · QoE at a Boundary (QoEaaB) · Smart Management

## 1 Introduction

With the big shift toward user-centric, service-centric, and transaction-centric interaction models, the role of users and their 'experience' along their journey of interactions with a provider has become an inseparable part of any design and assessment process of 'providing.' This vision has resulted in more complexity and also more variability that originates from the user's side and

Reza Farrahi Moghaddam · Mohamed Cheriet
Synchromedia Lab and CIRROD
École de technologie superéure (ETS), University of Quebec (UduQ)
E-mail: imriss@ieee.org
URL: http://ca.linkedin.com/in/rezafm

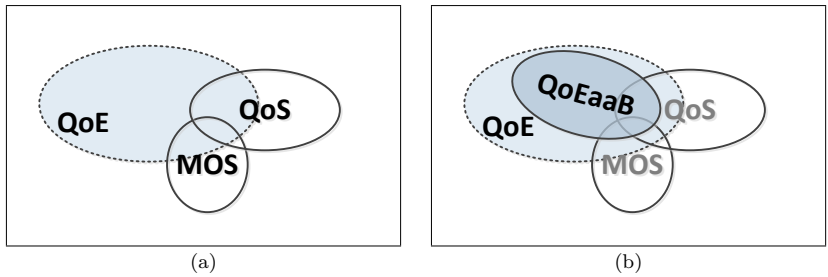

Fig. 1: a) An abstract illustration of overlap among the QoE, the QoS, and the MOS. b) The proposed concept of the QoEaaB, illustrated as an enabler toward exploring areas of the QoE that are not accessible to the QoS- and MOS-based approaches.

propagates to all other areas of service providing. Various measures, metrics, and indices have been put forward to enable quantification of these complex aspects, including but not limited to, Quality of Service (QoS), Quality of Experience (QoE), Quality of Resilience (QoR), Class of Service (CoS), and Grade of Service (GoS) (Stankiewicz et al 2011; ITU-T 2006, 2008, 2011). Although the number of metrics by itself shows the intensity of the associated complexity, the QoE metrics alone seems to inherit a high degree of challenge because of explicit involvement of the 'user' in its 'definition;' according to ITU-T Rec. P.10 (formally G.100), QoE may be defined as the overall acceptability[1] of an application or service, as perceived subjectively by the end user (ITU-T 2006, 2008).

Although this approach to QoE is by definition correct, it requires an interpretation in order to be become a quantifiable metric. Similar to many other subject-related metrics, the Mean Opinion Score (MOS) 'measurement'-based approach has been considered in quantifying the QoE (Stankiewicz et al 2011; Fiedler et al 2010). Although this seems to be a straightforward approach, the 'multivariate' nature of the problem, which mainly sources from the end user component itself, could make the final conclusions insignificant or even wrong if proper multivariate analyses are not considered (Jackson et al 2011). In contrast, the need to have a system-originated metric for the QoE, and also some conclusions drawn from the results of the MOS evaluations of the QoE in practice has led to developing models that try to directly relate the QoE to the QoS. We will discuss this aspect in more details in section 2.

---

[1] Overall acceptability may be also influenced by user expectations and context (ITU-T 2008).

In terms of the factors that influence the QoE, it has been argued that it is the small details in the user journey, in addition to the price and the service performance, that make the big differences in their experience (Bolton et al 2014; Baron et al 2010). Examples of these small variables could be i) emotions and ii) service recovery strategies, among others (Bolton et al 2014; Smith and Bolton 2002; Chan et al 2009). In this work, we argue that these small details are highly magnified at the boundaries of experience domains. Although all approaches to quantify QoE, including those that are based on MOS or QoS, are correct and shed light on many aspects related to the QoE, we believe that there are areas in the experience domains that could not be directly accessed using MOS/QoS approaches. In other words, we suggest to model experience as a collection of various experience 'perspectives,' which could form an ensemble to provide a comprehensive understanding of the 'experience.' Figure 1(a) provides an abstract illustration of the perspective-based approach in which the overlapping areas between the QoE and the QoS, for example, practically show a 'perspective' to the QoE from the QoS point of view. The area that has no overlap with QoS or MOS is of our interest.[2] In particular, the area shown by the QoE-at-a-Boundary (QoEaaB) in Figure 1(b) is the focus of this work. We argue that boundaries in general are the richest areas that generate a considerable amount of experience and therefore highly influence the QoE.

The paper is organized as follows. In Section 2, a discussion on relations between the QoE and the QoS, and the need to consider the QoEaaB is provided. In Section 3, a use case related to multi-tenant resource sharing is discussed and a new approach based on the notion of *deny of request* is proposed. Then, in Section 4, another use case concerning the challenge of over the top video services and their impact on the broadband experience is presented. In addition, the question of 'boundary' itself is considered, and various instances of this concept ranging from property, contract, content, to time are discussed from a QoE point of view in Section 2.1. Finally, in Section 6, the conclusions and some prospects for the future are summarized.

---

[2] Although it could be argued that the MOS approaches should be capable covering the whole QoE domain, it is possible that there are aspects of the QoE that has been never thought of by any past/current user or opinion-question designer.

## 2 The Definitions of the QoE and the QoE-at-a-Boundary (QoEaaB)

As mentioned in the introduction, the ITU's definition of the QoE requires additional interpretation in order to be quantified. A popular strategy to this end has been on developing a relation between the QoE and the QoS. However, the QoS definition by itself is a challenge to be addressed. The QoS could be defined based on non-functional requirements (NFRs) of a service (Bartolini et al 2013; Briones et al 2010; Ameller 2014). The NFRs can be associated with those requirements that specify the system properties related to performance, security, and reliability, among others.

It has been observed in many studies that the nature of a service itself should be also considered in its evaluation. For example, services provided prior to a transaction are essentially different from those occurred afterwards (Berry et al 2010). Moreover, it has been noticed that composability of the QoS is of great importance, especially when the complexity of the service orchestration and choreography increases (Briones et al 2010). In terms of the QoE, the influential factors could go beyond the system itself, and a minimal partitioning based on three influential factors of human, system, and context seems to be required (Reiter et al 2014).

Many approaches have been considered to make the QoE a function of the QoS. For example, a relation has been developed between QoE and QoS based on a hypothesis that there exists an exponential interdependency between 'parameters' of the quality of experience and those of service (the IQX hypothesis) (Fiedler et al 2010; Varela et al 2014). In Goran et al (2014), an exponential relation has been also developed between the QoE of the IPTV services and the associated QoS parameters, such as the QoS of the application, network, and physical link.

In general, as has been acknowledged in many researches mentioned above, the QoE has some important overlapping areas with the QoS. However, the overlapping area would not be a complete picture, and therefore the focus of this work is on the non-overlapping part. In particular, a subarea defined by the role of 'boundaries' on the QoE is the main target. When operating at the proximity of a boundary, changes in the degree of the proximity or events of crossing the boundary could have a great impact on the user experience, and therefore they should be explicitly considered in the QoE. In this work, this perspective of the QoE which considers the influence of boundaries is called the QoE-at-a-Boundary (QoEaaB). Identifying the boundaries would be the main challenge considering the fact that many of them would not have any 'tangible' counterparts. The following sections provide some instances of boundary, and also some more discussion on boundaries are listed and discussed.

2.1 The Devil is at the Boundaries

The first step toward a vision of the QoEaaB is a proper definition of boundary; a straightforward definition could simply be the boundary of 'things.' Although in the common sense, a thing is usually a 'property' that can be owned, we will see in the examples below that there are other possibilities, ranging from contract to content, which could impose boundaries that would drastically influence a user's experience, perception, and satisfaction. We start with two use cases: i) An example of the boundary of Service-level agreement (SLA) in a use case of multi-tenant resource sharing (Section 3) and ii) an example of boundary of content in a use case of Over-The-Top (OTT) multimedia services (Section 4). Other boundaries, and in particular the boundary of property will be discussed later on. The main focus would be on aspects that would be undiscoverable using other approaches to the QoE.

## 3 A Case Study of the Boundary of SLA: Multi-tenant Cloud Services

In terms of SLA, Various boundaries could be considered depending on the nature of the actual users involved. In this section, a case study of the multi-tenant scenarios is discussed. In these scenarios, each tenant actually represents a group, i.e., its users. Therefore, depending on the nature of a tenant's business, its users would show statistics that could have not been accounted for by the tenant when they set their service layer objectives (SLOs). Although in an isolated operation, where there is only one tenant being served, handling the risk associated with users variability of the tenant is the sole responsibility of the tenants themselves, the risk would propagate to other tenants in actual realistic multi-tenant cases. Although these events of service denial or degradation would be highly rare, and mostly show temporal correlation with "social events,"[3] the fear of negative experience on the users would result in reluctance of many potential tenants to join the multi-tenant ways of operation. In addition, it is worth noting that exponential growth events initiated from service requests of a tenant in the multi-tenant solution could show similar statistics of denial of service (DoS) attacks. However, they should not be classified as attacks because the request issuer has no intention to harm. We will discuss this point more at end of this section.

---

[3] Such as sport matches (Evens et al 2011) or the TV prime-time hours.

In general, multi-tenancy is of interest because of its potential in enabling i) resource sharing as well as ii) elasticity. The latter[4] can be seen as a practical exercise of the 'right to include' (see section 5.1). Tenants willingly participate in a 'community' in which they foresee that by providing other tenants with their prospective designated resources at times they do not require them, they can in return have access to other tenants' resources at the time they are in need of more resources.[5] Although an 'elastic' agreement actually seems to be a generic 'resource sharing' agreement, there is a critical difference between these two. In a 'generic' unregulated resource sharing operation, the participants could 'freely' use the resources without an explicit 'regulation' in place; this is usually the case of public resource provided to everybody. In these cases, the low level of usage compared to the high volume of available resources would make the generic resource sharing model acceptable for services provided by public bodies. In contrast, in the 'elastic' resource sharing operations, the resource pool is actually a potluck indirectly put in by all the participants, and on top of it there is a 'federal' regulator, i.e., the host of the multi-tenant operation who is responsible to optimally and fairly operate the resource allocation among the tenants.[6]

The advantage of multi-tenant operations would be considerable, especially when the tenants have some skewed features, such as different operation time zones. However, it would bring a great challenge considering the fact that the size of tenants, their inter-competition, and selfishness would be highly heterogeneous, and therefore opportunism could be strongly exploited. The fear of such a possibility would in turn discourage participation in such a federated resource sharing model. Therefore, complex and smart resource sharing strategies should be developed in order to attract the trust of the prospective tenants. Otherwise simple strategies, such as first come first served, would not be of any help. Two boundaries could be identified in multi-tenant solutions: i) the boundary of elastic resources that are horizontally shared among the tenants, and ii) the boundary of the service request entry. We will discuss these boundaries in more details in the following subsections.

---

[4] Elasticity usually stands on the top of the resource sharing, and is sometimes mistaken to be its equivalent.

[5] Although the resources are actually provisioned and owned by the host, the tenants could claim that they are virtually the resource contributors.

[6] It is worth nothing that there is another level of community management more commonly known as 'confederally-regulated' community (Brzinski et al 1999a,b; Hamilton 1787) in between the two cases of 'unregulated' and 'federally-regulated' communities. In the confederal communities, the requirement of having a federal body, i.e., the multi-tenant service host is relaxed at the cost of lower level of capability and optimality in regulating the interactions and resource sharing actions among the community members. Although federal models are usually preferred, the confederal models would be inevitable when the competition level among the members of the community goes beyond a threshold that would nullify any trustability in a federal body to fairly allocate the resources. We will explore more this aspect in the context of the ICT ecosystems in a future work.

3.1 Multi-tenant Cloud Services at the Boundary of Elastic Resources

Various approach have been explored to manage the elastic boundaries within the resources (Krebs et al 2014a,c,b). In short, the problem could be simplified to an optimization problem to set individual weights for every tenant in real time, which is then is used to divide resources among the tenants. Following the notation of (Krebs et al 2014b), the fitness function could be defined as a 'linear' sum of individual fitness functions of every tenant:

$$J[\mathbf{w}] := \sum_{i=1}^{n} J_i(w_i),\qquad(1)$$

where $\mathbf{w} = (w_i)_{i=1}^{n}$ is the tuple of tenants' weight, and $n$ is the number of participating tenants. For a typical tenant $i$, $w_i$ $(= w_i(t))$ at any time indicates the ratio of the resources that would be allocated to them ($0 \leq w_i \leq 1$).

As the first contribution, we introduce the 'deny of request' (DoR) process in the form of a random variable $X_i$ and a threshold probability $P_i$ which is defined based on $w_i$. With every service request from the tenant $i$, $r_{i,k}$, the random variable is inquired and the resulting number, denoted $x_i$, is placed against the threshold probability of that tenant at that time, i.e., $P_i$. Later on, we will define the value of $P_i$ based on the values of $(w_j)_j$. If the random number is less than $P_i$, that specific service request is denied:

$$\text{DoR}_{i,k} := \begin{cases} 1 & \text{if } (x_i < P_i) \ \& \ (P_i > P_{\min,i}), \\ 0 & \text{otherwise.} \end{cases}\qquad(2)$$

where the $\text{DoR}_{i,k}$ is the outcome of the deny-of-request process for the $k^{th}$ service request from tenant $i$. A $\text{DoR}_{i,k} = 1$ would result in dropping the associated service request $r_{i,k}$. The parameter $P_{\min,i}$ is a minimum deny probability required to activate the DoR process for the tenant $i$. The proposed DoR approach to multi-tenant resource sharing has an advantage over the traditional 'dividend among tenants' strategy based on the $w_i$ values; the DoR process partially allows the system serves spontaneous changes in the service request volumes of tenants even without updating their $(w_j)_j$ values. In contrast, the common dividend sharing approach would at times lock up a considerable amount of unused resources.

To define $P_i$, we assume an exponential relation between $P_i$ and $w_i$:

$$P_i := \exp\left\{\frac{1}{h_\mathcal{P}^2}\left(\frac{N_{i,\text{succ}}^{\text{sen}}}{\sum_j N_{j,\text{succ}}^{\text{sen}}} - h_\mathcal{P}^2 - w_i\right)\right\} - 1 \quad (3)$$

where $N_{j,\text{succ}}^{\text{sen}}$ is the number of 'accepted' service requests from tenant $j$ over a sensing time interval $\Delta t_{\text{sen}}$ (which is shorter that the time interval set to update the $w_j$ values) prior to the current time $t$, and $h_\mathcal{P}$ is a bandwidth associated to the DoR process. To define the $P_{\min,i}$, we identify the tenant with minimal $P$ value in the minority 'scope' of the tenant $i$, i.e., those with $w_j \leq w_i$:

$$j_i^* = \underset{\begin{cases} j : 1 \leq j \leq n, \\ w_j \leq w_i, j \neq i \end{cases}}{\operatorname{argmin}} P_j, \quad (4)$$

and a $P_{\min,i}$ value is designated which is negligible for the tenant $j_i^*$:

$$P_{\min,i} := P_{j_i^*} - \lambda_{\min}\left|P_{j_i^*}\right|, \quad (5)$$

where $\lambda_{\min}$ ($0 \leq \lambda_{\min} \ll 1$) is a ratio that defines the level of negligibility. Also, it is assumed that $h_\mathcal{P} < \sqrt{w_{j_i^*}}$, $\forall i$, at all time for the purpose of stability. In other words, it is assumed that a constraint in the form of $w_j > h_\mathcal{P}^2$, $\forall j$, is always applied in the procedure of determining $(w_j)_j$ described below. At extreme case of having very low volume of incoming service requests from all tenants in the minority scope of the tenant $i$, the $P_{j_i^*}$ value converges to a negative value of $\exp\left\{-w_{j_i^*}/h_\mathcal{P}^2 - 1\right\}$ - 1. This means that the DoR process

Quality of Experience at a Boundary (QoEaaB) 9could be activated for the tenant $i$, simply if

$$\exp\left\{\frac{-w_{j_i^*}}{h_{\mathcal{P}}^2} - 1\right\} - 1 < \exp\left\{\frac{1}{h_{\mathcal{P}}^2}\left(\frac{N_{i,\text{succ}}^{\text{sen}}}{\sum_j N_{j,\text{succ}}^{\text{sen}}} - h_{\mathcal{P}}^2 - w_i\right)\right\} - 1,$$

or,

$$N_{i,\text{succ}}^{\text{sen}} > (w_i - w_{j_i^*}) \sum_j N_{j,\text{succ}}^{\text{sen}}.$$

Calculation of the $w_i$ shares is essential in practicing the proposed DoR process. Following Equation (1) and (Krebs et al 2014b), the cost functions of an individual tenant cost function could be defined as:

$$J_i(w_i) := H_i(l_i) p_i(l_i) v_i(w_i), \tag{6}$$

where $H_i(l_i)$, $p_i(l_i)$, and $v_i(w_i)$, and $l_i = N_{i,\text{succ}}^{\text{upd}}$ are the heaviness, penalty and violation functions, and the workload of the tenant $i$, respectively. The workload $N_{i,\text{succ}}^{\text{upd}}$ is defined the number of accepted service requests from tenant $i$ over an 'updating' time interval $\Delta t_{\text{upd}}$ ($\Delta t_{\text{upd}} > \Delta t_{\text{sen}}$) prior to the current time $t$. The heaviness $H_i$ at any time $t$ is defined as the ratio of the workload of the tenant $i$ at $t$ to the averaged workload of all tenants at $t$:

$$H_i := l_i n / \sum_j l_j = N_{i,\text{succ}}^{\text{upd}} n \Big/ \sum_j N_{j,\text{succ}}^{\text{upd}}.$$

The penalty function $p_i(l_i)$ is defined as a flipped sigmoid function to account for excess of $l_i$ compared to the quota of tenant $i$, i.e., $q_i$. This function can be expressed as a function of the parameter $H_{i,i}^0$, i.e., the ratio of $l_i$ to $q_i$:

$$H_{i,i}^0 := l_i/q_i,$$
$$p_i(l_i) := 1 / \left(\exp\left\{\left(N_{i,\text{succ}}^{\text{upd}}/q_i - 1\right)/h_p\right\} + 1\right).$$

Finally, the violation function is defined as an exponential function of difference between the response time and the guaranteed response time value $g_i$ of the tenant:

$$v_i(w_i) = \begin{cases} \exp\left\{\left(\delta\left(w_i^-\right)/g_i - 1\right)/h_v\right\}, & \delta\left(w_i^-\right) > g_i, \\ 1, & \delta\left(w_i^-\right) \leq g_i, \end{cases}$$

where $\delta\left(w_i^-\right)$ is the mean measured delay in the response to the tenant $i$ requests with current value of $w_i$ (at $t^- = t - \epsilon$), and $h_v$ is the bandwidth parameter shared among all tenants. Please note that $w_i$ is a constant on the interval $[t - \Delta t_{\text{upd}}, t)$.

Below, we propose several other enhancements, in addition to the DoR process, for the multi-tenant fitness function. The combination of these enhancements is summarized in a proposed fitness function in Equation (7) below.

1. **Nonlinear combination of individual fitness functions**: when tenants' size varies in orders of magnitude with respect to each other, the fitness of some of the tenants would act as heavy-weight outliers, which would then compromise the service for other tenants. We will consider nonlinear transforms (Breaban and Luchian 2013; Zhou and Barner 2013; Cunningham and Ghahramani 2014) to regularize these outliers without explicitly removing them from the calculations.

2. **Addition of memory to the system (reputation of individual tenants)**: There is a weakness to a memory-less fitness function since it would repeat the same mistake providing the fact that the same faulty workload configuration appears. By including a minimal memory, such as that of a Markov model,[7] the optimizer could learn from the incidents in the past, and then it could reduce the probability of compromising the guarantee of those tenants that work within their quota in future incidents. This could be also incorporated in the form of a reputation factor for each tenant that keeps track of their volatility behaviors. In addition, the memory-enabled models could be used to develop predictive models, as discussed in the next item.

3. **Addition of short-term behavior prediction**: even with the most accurate fitness model, the response time to an incident of exponential increase in the workload of a tenant would suffer some delays. These delays are mainly caused by the requirement to 'see' the 'violation' in the service in order to apply weight recalculation. By adding predictive models, which are inferred from the possibility of an incident in highly short intervals based on temporal trends, the optimizer could react and accordingly adjust the weight prior to the event in order to avoid any 'glitch' in the service to other tenants. Memory-based and pattern-based models could be used for such a purpose.

4. **Load-adapted guarantee values**: the quota of a tenant and its guarantee, i.e., $q_i$ and $g_i$, are practically independent control parameters of the optimizer. Although the ratio of $l_i$ to $q_i$ was used in the penalty function, the ratio of $l_i$ to all workload quotas would be another important parameter in managing possible high levels of variability.[8] Therefore, there would be three heaviness variables: the first two are the previously defined $H_i$ and $H_{i,i}^0$, which measure the ratio of the $l_i$ to the averaged workload and its quota, respectively. The third variable is denoted by $H_i^0$, and is defined as the ratio of the workload to the averaged quotas:

$$H_i^0 := l_i n \Big/ \sum_j q_j.$$

---

[7] For example, Hidden Markov Models (HMM) or more complex Hidden semi-Markov Models (HSMM) (Yu 2010).

The direct consequence of introducing $H_i^0$ would be the requirement to have a load-adaptive $g_i$. In other words, $g_i$ should be compromised when there is an increase in either $H_{i,i}^0$ or $H_i^0$. The reference values for $H_{i,i}^0$ and $H_i^0$ would be 1 and $q_i n / \sum_j q_j$, respectively.

5. **Addition of 'binary' inter-tenant fitness functions**: in a heterogeneous multi-tenant community, many tenant 'pairs' could be identified based on common features of paired members (two competitive tenants would easily form a pair). By defining binary inter-tenant fitness functions for every or some of these pairs, the response of the service optimizer in providing a fair and optimized resource allocation could be improved. In particular, many isolation measures could be implicitly implemented using these binary functions. Furthermore, addition of binary terms would help to convert the main fitness function into a standard form that could be rapidly solved using global optimizer, such as the graph cut optimizer (Boykov and Kolmogorov 2004), without any requirement to use iterative numerical solvers.

Hereupon, we propose a new fitness function:

$$J[\mathbf{w}] := \sum_{i=1}^n J_i(w_i) + \sum_{i,j=1}^n C_{i,j} J_{i,j}(w_i, w_j), \qquad (7)$$

where $J_{i,j}$ represents the inter-tenant fitness function of tenants $i$ and $j$ of a typical pair tuple $(i, j)$, and $C_{i,j}$ is the correlation between $i$ and $j$. A value of zero for $C_{i,j}$ would mean that the pair $(i,j)$ does not exist. The intertenant terms would explicitly depend on the heaviness variables. For a pair of completely correlated tenants, the following $J_{i,j}$ is proposed:

$$J_{i,j} = \prod_{k \in \{i,j\}} \left( \frac{R\left(w_k - H_k^0\right)}{H_k^0} \right) + 0 \left\| \frac{w_i}{w_j} - \frac{H_{i,i}^0}{H_{j,j}^0} \right\|, \tag{8}$$

where $R(\cdot)$ is the ramp function:

$$R(\eta) = \begin{cases} \eta, \ \eta \geq 0 \\ 0, \ \eta < 0 \end{cases}.$$

---

[8] Although a model based on $H_i$ and $H_{i,i}^0$ would be robust for quasi-static competitive tenants, the 'normal' behavior would change from time to time depending on the averaged workload experienced. These changes would not be acceptable in the multi-tenant operations in which each tenant implicitly puts in a portion of the resources in the federated community. Therefore, it is expected that the ratio of the workload to the averaged quota of all tenants should be considered in the modeling.

The second term has a coefficient of zero assuming 'total' competitiveness between the members of the pair. However, in milder cases, where there are less levels of competition, this coefficient can be defined as a function of $C_{i,j}$.

The proposed model in equation (7) horizontally explores the boundary of SLA in multi-tenant resource sharing operations. In future, we shall explore its potentials.

3.2 Boundary of the Service Request Entry

The second boundary in terms of our case study is the service request entry. In contrast to common sense understanding of this boundary, we would like to explore possible opportunities to create better experience for the tenants. Figure 2(a) shows the standard picture of a multi-tenant service solution. The actual configuration could be different, but the main components are the same. As mentioned earlier, the rare events of exponential growth of tenant service requests could result in default of the whole multi-tenant community. Although the screening firewalls and the admission gate components are planned to disable intentional attacks and to regulate request streams, in extreme cases, the information processing of these protective components themselves could exhaust their available resources, and therefore lead to an unpreventable default at the 'wall.' Even when the abnormal tenant is identified, there is another challenge because the stream would be still directed to the host by the transport network.

The approach proposed in Figure 2(b), illustrates smart transport network that could transit the requests and their fragments[9] not only based on their destination address, but also by considering their 'source' features in the decisions.[10] This solution interestingly resembles the Software Defined Network (SDN) concept. However, in contrast to many other applications, controller-based solutions would not be applicable here at the Internet scales because of independence, heterogeneity, and highly-distributed nature of the transport network.[11] A preferred approach to a smart transport network would be based on 'smart protocols' and 'trusted messaging mechanisms' that allow rapid reaction to abnormalities even before full deployment of the mitigating configuration across all transport network. A non-local Traffic Regulator Hub (TRH) in this solution in considered to handle and regulate redirected requests. Figure 3 shows an illustrative and simplified response of the smart transport network to a redirected request warranted on an exponentially growing request stream of a particular tenant.

---

[9] For example, packets in the IP networks.

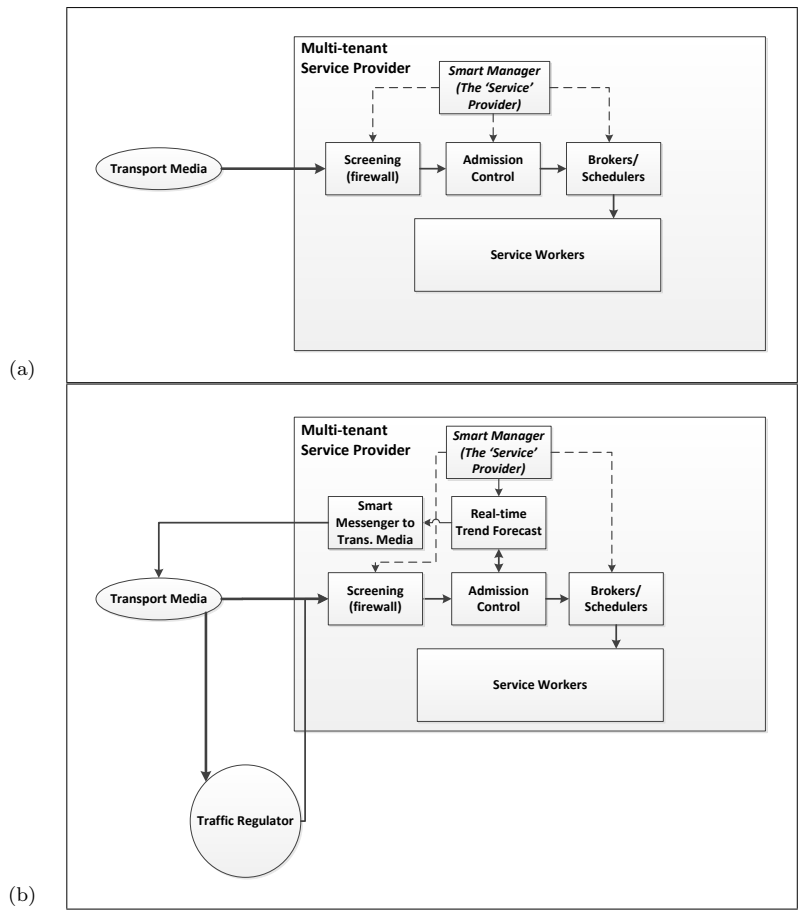

(a)

(b)

Fig. 2: a) A schematic illustration of fundamental components of a multi-tenant service solution. b) The proposed solution with an additional messenger unit to the transport network, and also a non-local traffic regulator unit.

---

[10] The criteria do not need to be limited to addresses and many other features of the tenant, and their authorized sources could be used in the decision making.

Although mechanisms such as Remote Triggered Black Hole (RTBH) (Turk 2004) and source-based remotely-triggered blackholing (S/RTBH) (Kumari and McPherson 2009),[12] have been practiced to mitigate distributed denial of service (DDoS) attacks in the form of blackholing the whole traffic directed to a victim destination or the portion of the attacker traffic sourced from a specific location, the proposed mechanism here can be seen to be different from various aspects. Firstly, the exponential growth in the service request cannot be classified as an attack, and therefore it should not be handled with the attack mitigation approaches. In other words, every service request from the abnormal tenant or from the other tenants should be treated as a legitimate request. The proposed TRH provides a second chance to the requests from the exponentially growing stream. Secondly, any harm to the host access would substantially affect its other tenants, and therefore it would reduce the acceptability of multi-tenant concept. The proposed mechanism would provide a way to split the streams aimed to the multi-tenant host, and protect the quota-abiding streams. Thirdly, the service requests from a tenant could be less source-specific and more 'task'-specific. Therefore, a more complex language at the protocol level is required in order to code the specifications of the sender rather than its address, and then use the coded information in order to make smarter decision at the routing nodes. It is also worth noting that decoy routers and servers have been proposed in order to redirect the traffic of an attacker off the victim servers (Okada et al 2014). In contrast, the proposed TRH concept can be seen as another multi-tenant service solution that serves many multi-tenant hosts at a hierarchically higher level. In other words, the same beneficiary advantages of the multi-tenant business model could become the motivation for a multi-tenant TRH business that allows sharing the TRH's Total Cost of Ownership (TCO) among its clients (which are actually other multi-tenant businesses). This could be of great interest considering the rareness and possible uncorrelated nature of the exponential growth events across different multi-tenant host. Moreover, the fact that the TRH service would be placed outside the immediate access networks of the providers would further protect them even in the case of the failure of the TRH service itself.

---

[11] The transport network could be imagined in the form of NSPs at different tiers among other networks.
[12] Based on Unicast RPF Loose Check (Greene and Jarvis 2001; Savola and Baker 2004).

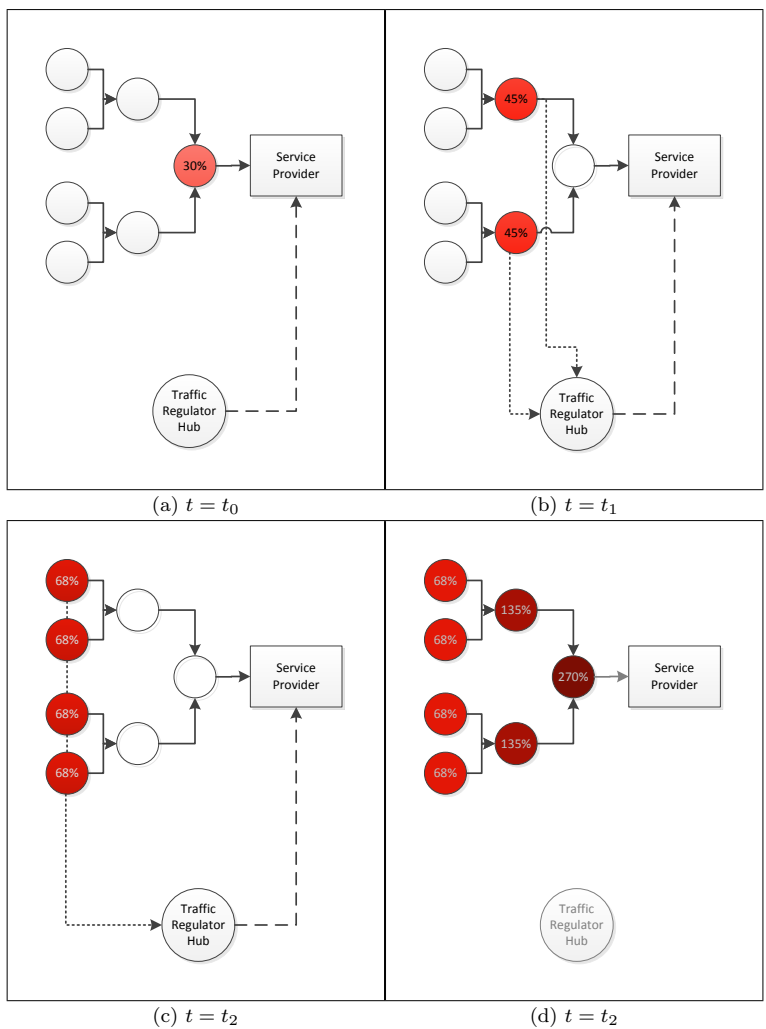

Fig. 3: An illustrative example of the real-time response of the messaging mechanism to an exponential growth in the service requests from a tenant in a multi-tenant scenario. There are two access transport network (NSP networks) that connect the multi-tenant host to the tenants and the tenant-authorized request sources. a) At $t = t_0$, the exponential trend (tripled after each time interval) is discovered at a 30% load at the screening component of the host. b) At $t = t_1$, the redirected message has propagated to the first switch of the networks. A bearable load of 45% is noticed on the transport network' switches. These nodes redirect this load to the non-local TRH. The regulator hub would direct a safe portion of the traffic to the service host. c) At $t = t_2$, the total load targeted at the exploding tenant could be a terminating load of 270% of the capacity of the screening switch. However, with propagation of the message within the transport media, the load on the individual nodes at the 'edge' of the redirecting message 'wave' would be still safe and around 68%. The redirected traffic is regulated at the TRH as planned. d) For the purpose of comparison, the load on the screening node would be 270% at $t = t_2$, if we do not consider the redirected messaging mechanism and the TRH. This is a certain default for the screening component and therefore the whole multi-tenant solution.

3.3 Multi-tenancy approach to virtualized radio access

Although the focus of the case study of this section has been on compute services, a similar approach can be used as a multi-tenant solution for radio bandwidth and access sharing. In particular, and considering developments in virtualized base transceiver stations (BTSs) of cells of various sizes (Costa-Pérez et al 2013; Ghebretensaé; et al 2010), it is possible to plan a multi-tenancy service that serves several Telco operators. Although the operators would have still their own access networks to the BTSs, they will leverage on the benefits of radio resource sharing that would benefit all participants. Moreover, aggregation of the operators' base stations in the form of the multi-tenant host's BTSs would reduce challenges and also costs associated with interference, especially in dense urban areas. However, the challenge would be, considering high level of competition, development of a highly trustable resource management to attract the operators. In future work, an extension of the multi-tenancy management proposed in section 3.1 will be considered for the case of virtualized radio access networks.

**4 The Case of a Netflix-like Video Provider/Distributor and ISPs**

There has been a big discussion and debate around the Over-The-Top Online Video Services (OTT OVS).[13] In particular, handling the video traffic through the chain of content distributors, content-delivery networks (CDNs), and retail ISPs has attracted a considerable attention (Frieden 2014; Krämer et al 2013). Although there have been concerns on whether the associated observed congestion is really related to resource shortage (Frieden 2014), we assume the validity of the shortage-driven congestion in this section.

---

[13] A few examples of OTT OVS, especially OTT Subscription Video on Demand (SVoD) services, are Netflix, Shomi, Crave TV, and HBO Go, among others. In the text, we may informally use Netflix as the example. However, the discussions generally apply to all OTT OVS.

The resource management in overloaded networks is a challenge, and would require revisiting over-promising SLAs traditionally offered to the end users, in the form of imposing quota on the delivered bits rather than the bandwidth alone. In addition to these static approaches, much more can be achieved in the form of collaboration between the content providers and ISPs toward increasing the QoE of the end user even in bandwidth-limited networks.

An example of such collaboration could be at the 'boundary' of the video streams themselves. During the prime time hours, in particular, numerous streams originated from the content provider or their CDNs could be aggregated over short time intervals of a few seconds, for example 5 seconds. We propose a technique called the 'micro-registration' of streams to enable considerable reduction in the required total bandwidth in the exchange of a few seconds launching-delay for the end users. The aggregated streams can be seen as virtual penetration of the CDNs within the access networks. The streams could be imagined as 'cars' of a train on which new end users are boarded depending on their service request timing.[14] Although the proposed registration time of 5 seconds is much bigger than the suggested 2 seconds limit of the 'channel zapping' delay (Lee et al 2010), we could argue that in our case the user has targeted a specific content and therefore they are not in a random 'channel browsing' mode. This makes the extended delay bearable by the user, especially if they are aware of it as part of their service. At the same time, the proposed dynamic, temporal, and smart aggregation of the content streams would provide an implicit way to push VoD operation toward broadcasting, and in turn would allow semi-multicast approaches to be used in order to handle the bundle of streams (Ramos 2013a,b; Deering 1991). Table 1 provides a summary of benefits of the proposed micro-registration approach on the actual bandwidth of every viewer during the prime time hours in a case of a metro area of 10,000 individual access points served using a connection of 2,000 Mbps. As can be seen, the bandwidth is improved by a factor of 27.78. It is also worth mentioning that other alternatives to our micro-registration approach have been considered including caching at the edge (Wan et al 2014; Kuenzer et al 2013; Frangoudis et al 2014; Abujoda and Papadimitriou 2015).[15]

---

[14] It shares a metaphorical analogy with the transportation sector, in which the public transport scenario (i.e., the train) reduces the energy consumption and the natural (environmental) footprint while increasing the fluidity of the traffic and reducing the commuting time compared to the individual transport scenario (i.e., personal cars). Similar to the transportation section, all these benefits depend on development of a mutual collaboration among all the actors involved including the landlords (peering and last-mile ISPs) and train operators (Netflix).

| Case | Total bandwidth to metro area | Number of active streams | Actual observable bandwidth |
|---|---|---|---|
| Without Micro-Registration | 2,000 Mbps | 10,000 | 0.2 Mbps |
| With Micro-Registration | 2,000 Mbps | 360 | 5.56 Mbps |

Table 1: Comparison between unregulated and micro-registration approaches to OTT OVS utilization peak in the prime time hours. It is assumed that the viewers starts their sessions during an interval of 30 minutes after 6PM. The micro-registration approach was able to achieve a delivered bandwidth of 5.56 Mbps which is higher that the recommended bandwidth for HD videos, i.e., 5 Mbps.

Beyond the SVoD services, in which content has its traditional meaning, new challenges are on the horizon with the rise of digital social networks. Exponential growth of the number of connected individuals and their associated devices to billions (Ericsson Inc 2011) would result in exponential increase in the number of events such as personal videos that become micro-viral. These content making/sharing episodes are: i) more variant, ii) more 'localized,' and iii) more ephemeral, and their nature makes them highly dependent on the seamless integration of social and Telco networks. This would push the Telcos to provide Over the Top (OTT) as a service, possibly in the form of Over the top of Telco Clouds (OTT-TC or $OT^3C$). In future work, an approach will be proposed based on smart integration and also a generalization of the micro-registration technique in order to reduce and contain the energy consumption and footprint of personal content micro-broadcasting.

Another approach to address limited bandwidth of networks could be built on the boundary of content. In contrast to traditional understanding of a content that was more about its physical carrier than the content itself, the new ways of access enable more general representations that could go beyond limits of the physical world. One advantage would be to represent an online content in a non-local manner over a vast geographical area. The non-local representation is beyond the cloud-based representation, because in the later although the content could migrate across a vast area, it is always attached to one of those 'locations' at any time. The non-local representations, which have been exploited before in the context of point-to-point content sharing, allows a content to be partially present at various locations at the same time. The non-local approach would relax the constraints to deliver the content from the provider/CDN to the end user per service request. Although there will be some questions in terms of copyrighted content, development in terms of content encryption and handling could provide practical solutions (Cieply 2014; Silverston et al 2009). Moreover, ever increasing volume of distributor-originated content would practically lead to a copyright agnostic future. The concept of non-local representation of data could be further pushed toward other concepts such as data 'immersion' and 'ICT as a part of life.'

---
[15] The Streaming Video Alliance and Open Cache: http://www.streamingvideoalliance.org/

Finally, as mentioned in section 5.1, there are other boundaries of content including the boundary related to encoding/decoding (Grois et al 2013; Shafique and Henkel 2014). This boundary could be also tapped in to increase the QoE even when the congestion is high in the network.

## 5 Other Cases of Boundaries for the QoEaaB

5.1 Boundary of Property

Before discussing the QoEaaB related to properties, it is worth mentioning that although ownership of a property is traditionally considered an advantage, with new categories of property in the horizon, especially considering intangible virtual properties, the liability and responsibility costs associated with ownership of all properties owned by an entity (or a person) have become a major burden.

The QoEaaB of properties have two facets. The first facet is directed toward isolation and separation of users' property, while the second one focuses on the 'inclusion' of a user in their provider's properties. Although our focus will be on the second facet, we would like to mention an example of the first one, in which compromise of the identify property (identify theft) of a user has been handled with a zero-liability policy in order to presumably preserve the QoE score (Grimmelmann 2010). However, the user's dissatisfaction reported in Grimmelmann (2010) shows that crossing the boundary of the user's property could have a considerable impact on the QoE even when it is handled with the best practices.

A common definition of a 'property' is the well-known 'bundle of relations' (Chang 2014), which is one step ahead of the 'bundle-of-rights' in the sense that it considers not only the owner but also all other parties involved. From the rights point of view, the important 'right to exclude' has been suggested to be accompanied with the 'right to include,' which would promote cooperation in various forms: informal, contractual, and proprietary inclusion (Kelly 2014). It seems that the right to include has a great potential for the service providers, especially in increasing their interaction and connectivity with the users as well as building long-term relations. In another approach to the definition of property, the 'property matrix' has been introduced as a new definition of property considering two dimensions of property: i) the descriptive dimension which requires a thing to give rise to a right[16] and ii) the normative dimension[17] (Morales 2013). In another perspective, the property is defined as the law of things, in the form of an architectural theory based on information costs with the advantage of using modularity to manage the associated complexity (Smith 2012, 2014). In this theory, the things were considered as modules, and instead of a bundle, a 'structure' of relations among the features is proposed for the definition of a property.

In general, the boundary can be seen as the *in rem* aspect of a property. Intangible properties such as information properties, face more challenges in finding an on/off proxy for violation of rights, similar to the case of water as property (Smith 2012). Although behavior at the proximity of the boundary of a property is considered here, there is another type of boundary that can be related to the property concept: The definition of property itself has boundaries, especially when it is put against the definition of tort. These boundaries would involve both the property rules and the liability rules as enforcers (Miceli 2013).

As mentioned before, the uncertainty and confusion in handling boundary of properties is deepened in the case of intangible properties. Intangible properties are of great interest in ICT and other dematerialization industries (Farrahi Moghaddam et al 2014a). It is expected that, in a move toward a sustainable future, ICT-enabled solutions would replace a considerable volume of services provided by other industries in order to reduce resource consump-tion, GHG emissions, and environmental footprint (Farrahi Moghaddam et al 2014).

---

[16] For example: 1. Rights of a person over a given thing, 2. Rights *in rem* (in contrast to rights *in personam*), and 3. Attachment to the thing (Morales 2013).

[17] 1. The right to exclude, 2. A bundle of rights, 3. Rights of autonomy, and 4. Rights of value enhancement (Morales 2013).

The scale of data and information collected by the service providers shows a great potential to 'process' the data property a service provider augments in such a way the users could be 'included' across the boundary of that data property. The impact on the QoE would be exponential, while the provider would also enjoy the insights in return for the costs of anonymizing and other associated data processing tasks (Bhimani and Willcocks 2014). This information transparency approach would also help in deterring opportunism which is a side effect of the information era and a result of the imbalanced availability of data (Steinle et al 2014).

Another example that highlights the importance of the QoEaaB is the case of artistic work properties, such as movies, songs, and books, and multimedia in general. Traditionally, artistic properties have been associated and practically mistaken with their physical carrier, for example DVDs. However, with the move toward online media, the media carrier's role is becoming less visible to the user, and at the same time its associated costs have been reduced thanks to resource-sharing aggregators, such as virtualization and cloud-based models (Farrahi Moghaddam 2014). With the rise of Over-The-Top (OTT) multimedia service providers (Farrahi Moghaddam et al 2014b), there is a big challenge and also opportunity to contain and reduce the impacts associated to these new business models that would deeply affect many industries including TV, ISP, and advertisement. This has brought up the question of how to handle the boundaries of an artistic work without taking into account its carrier. We can observe that there is a great potential for novel ways of inclusion of user inside the artistic property that practically prevent forcing users to make hard decisions in terms of owning an 'instance' of the intangible work. An example of such inclusive boundary crossing could be special, limited preview or view licenses delivered by OTT providers.

5.2 Boundary of Contract (Boundary of User Decision)

The next item on our list of 'things' is the contract. It worth mentioning that the historical justifications for the necessity of contracts would be no longer valid in many cases in near future, and therefore the whole concept of contract would require a complete reconsideration. That said, hereinafter we assume that the contract is a necessary part of any interaction between a provider and a user. Further on, it will be argued that the boundary of a contract would play a significant role in the QoE of the user.

A contract is usually a documentation of a hard decision making. Although this may be vital in order to stabilize operation on the provider's side, many factors need to be considered. The first factor is the user itself. Although it is assumed that a user, similar to all other actors, is a 'selfish' player, there is no guarantee that the user's decision is always 'optimal' (Staub-Kaminski et al 2014). A typical end user has very limited analytic resources, and therefore they highly depend on their social 'connections' in order to make a decision.[18] The non-optimality of the decisions would result in dissatisfaction and a poor experience for the user each time they approach the boundary of the contract associated with those decisions. Therefore, there is a potential that novel management techniques of the boundary of a contract bring in great opportunities to increase the QoEaaB and the QoE in general without scrapping the concept of the contract itself. One way of tackling this problem would be providing 'on-the-fly' options to modify a contract.[19] However, it is more desirable to provide those options the user has not even asked for or thought of. In other words, the ability to 'observe' the degree of the user's proximity to the boundary, and thereupon offering in-kind or fair options not only increases user's satisfaction, but also helps in increasing their trust and loyalty in the provider.

In another direction, it has been pointed out that the contract concept is a better interaction 'container' than that of the license concept when formalizing relations (Patterson 2012). In the case of intangible properties in particular such as those of ICT,[20] a license may fail to transfer the conditions required by the provider, and especially it may fail in obtaining the consent of the receiver on those conditions.[21] This would result in the complexity and challenges in designing and enforcing licenses, especially in terms of what crossing their boundary would mean. A more properly placed boundary, and more importantly, a smart way of handling the circumstances at the proximity/crossing of the boundary would provide a great chance in improving the QoE.

Finally, it is worth mentioning that the traditional definition of a contract would not be proper in new visions to human ecosystems, such as that of

---

[18] This could be actually identified as a drive for many social networks.
[19] Even, temporarily for a short period of time.
[20] For example, open-source software, and artistic works, among others.
[21] In addition, the extra information cost resulted from the added restrictions, which would be finally translated into rights, would in turn increase the processing cost of every potential acquirer even if they do not have any commitments to make a purchase (Patterson 2012).

Zero Marginal Cost Society (Rifkin 2014), in which the capital cost of a service provider would become negligible, especially for intangible properties. In those cases, smart service providing models could create opportunities at the 'boundary of the service' itself.

5.3 Boundary of SLA

In many cases, the non-functional requirements, mentioned in Section 3, are actually listed in the SLA instead of the contract. Therefore, the boundary of an SLA would have a critical role in defining the experience of an user. Formally, an SLA governs management of non-functional parameters based on some restrictions aiming to optimize a cost function. It can be argued that the 'cost function' itself is the main factor that determines the boundaries of an SLA and also the behavior when crossing those boundaries.

The first question in terms of designing an SLA cost function is who actually benefits from the service. There are two cases: i) it is an end user and ii) it is an intermediate 'tenant' that provides a secondary service to other users. In the former case, the QoE should be defined with respect to 'an ensemble' of end users, and therefore the traditional notion of 'average' would not be proper. In addition, the experience would have a 'nonlinear spectrum' of intensity, ranging from 'delay in delivery' to 'no delivery at all.' The ensemble view on these experiences should be well defined, considering possible conflicts between the 'individual' and 'society' perspectives toward the experience. In other words, the 'rare' but 'abnormal' experiences require a special attention in the cost function in order to protect the experience of the 'individuals' from being 'shadowed' by the 'frequent' normal experience events associated with their 'society.' This brings up the notation of the 'boundary of an ensemble.' To be more precise, this requires another level of service in order to handle the requests to be included in the ensemble. The experience of this secondary-level service could be very extreme, and therefore could have a drastic impact on the out-of-the-ensemble users' impression of the provider. Although imposing a boundary for the ensemble would be critical in preserving the quality for the insiders, smart handling mechanisms of the newcomers' inclusion-requests are necessary. These smart mechanisms would be probably dynamic, 'temporal,' and highly dependent on the elasticity of the ensemble and that of the provider's resources.

The role of SLA is more critical when resource sharing and virtualization approaches are in place. The semi-intangible nature of the virtualized resources requires proper handling of the QoE and quality of isolation metrics. In Hobfeld et al (2012) and Ksentini et al (2014), various challenges from migrating Online Video Services (OVS) to cloud environments have been identified. In addition to many factors, such as latency, management, and multi-party communication, the SLA and pricing are required to provide flexibility in terms of guarantee of video quality versus price. This is in line with the QoE at the boundary of 'delivered video-quality' as a property.

As mentioned in Section 3, in contrast to the case of 'ensemble of end users' where the boundary of the ensemble is the most influential area, in the case of multi-tenancy it is the boundary of the semi-intangible resources assigned to a tenant that determines the quality of experience and isolation. Dynamicity and shrinkability of these resources, which somehow originates from the lack of a governing contract, would make the boundary highly variable, and therefore this could increase the frequency of the boundary crossing events and their associated negative experience and costs. 'Grading' protocol or standard of virtualized resources is required in order to reduce variability while maximizing utilization of the associated [tangible] infrastructure resources.

5.4 Boundary of Context

The context has been well identified as a key factor in determining the QoE (De Moor et al 2010), and boundary of different contexts would be the most challenging area to handle because of possible confusion in identifying the dominant context or probably ignoring the direction of the 'transition' across the boundary of adjoined contexts. We will explore this instance in more details in future work.

5.5 Boundary of Content

Although the concept of content was earlier addressed from the perspective of property in Section 4, it appears crucial to discuss it separately here. In general, information, data, and many other forms of intangible properties could make a big difference in the QoE at their boundaries. A popular example is the artistic content, such as movies.

A fast growing content-related service sector is associated with online video delivery. Although the business models have a wide spectrum from paid IPTV/OTT models (Goran et al 2014; Farrahi Moghaddam et al 2014b) to ad-based models such as YouTube, all models are gradually converging to use a 'generic' broadband media, such as that of the Internet, smart grid or smart house telemetry backhaul, and even carrier network (Gomez-Cuba et al 2014, 2013), in delivering multimedia with constantly-increasing video qualities. This move has brought a common set of challenges to all these models. Various boundaries could be identified and then tapped on in order to improve the QoE, and at the same time reduce the associated natural (environmental) footprint. Some of these boundaries are directly related to the content itself: i) the boundary of encoding/decoding, ii) that of encrypting/decrypting, iii) the boundary of observable video quality, and iv) the boundary of centralized vs. distributed content storage. Other associated boundaries are human-computer interface (Farrahi Moghaddam et al 2014) and transport protocols (Hobfeld et al 2014; Rufini 2014). In particular, micro-registration of streams, which has been discussed in greater details in Section 4, can mitigate the negative impacts of online video services peak utilization during the prime time in the evenings.

In terms of protocols, there is a big concern that their designs suffer from a high degree of inefficiency that would translate in degradation in the QoS and QoE even when the 'physical' transport media is capable of delivering what is beyond the protocol-achievable bandwidth (Hobfeld et al 2014; Rufini 2014). It has been shown that simple management mechanisms added to the standard TCP protocol, for example, could increase the performance toward the 'dark fiber' limit (Rufini 2014). However, it can be argued that such mechanisms should be standardized and officially imported into protocols in order to extend their application to all instances. Otherwise, there would be a chance that a promising protocol is abandoned because of the low performance of its primitive form.

5.6 Boundary of Time

Finally, we would like to mention that boundary of time intervals could have a significant influence on the perceived QoE. Usually, these boundaries are much smaller than the time intervals themselves. However, the associated negative memories could be much more resilient than positive memories at other mo-ments (Moscovitch et al 2011). One day-to-day example is the rush hour traffic congestion for commuters. Another example could be the prime time hours in the evening for the TV/video users. A final example could be peak hours in electricity-scarce regions that would result in blackouts as an extreme experience.

A dimension associated with the boundary of the time intervals is the service providing 'journey' itself. Usually, such a journey consists of several sub-services that should be chained together to build the main requested service. All the required 'chaining' actions, not the chained services themselves, could impose extra processing overhead to the user, and therefore could lead to less trust. In contrast, practices which take the user from the starting point and automatically move them across the intervals and chained links in the form of a service omnibus would greatly improve the QoE and also the rate of success per request.

5.7 Monetization of the QoE

In contrast to the QoS that can be directly referred to in the SLAs and then be monetized, the QoE features could not be processed in the same way unless they are explicitly integrated or reflected in the QoS. Beyond that, the QoE could provide an implicit mechanism to induce loyalty of users to the provider, increase the user base, and improve the internal efficiency that make the business more profitable. The QoEaaB can be seen as an instrument to discover and identify innovative and user-involved actions that would provide a user with the peace of mind and at the same time would offer the provider with unique innovations that may 'differentiate' them from the rivals.

6 Conclusions and Future Prospects

Quality of experience has attracted many researchers considering convergence to a parity among providers in terms of many other metrics. In this work, a new approach to the QoE has been presented in order to provide a pathway for exploring unseen and novel opportunities to create high-quality experience for the users. The main aspect of this approach is the 'at a boundary' concept that results in the QoE at a boundary notion. As a case study, two boundaries associated with the multi-tenant solutions have been explored, and some

Quality of Experience at a Boundary (QoEaaB) 27

fitness functions and configurations have been proposed in order to provide better experience to the tenants. Also a use case of other-the-top online video services is discussed along the boundary of the video streams. In addition, various boundaries were examined and discussed, especially the boundary of property, contract, SLA, context, content, and time. In future work, each of these aforementioned boundaries will be considered as an opportunity to improve the QoE, and proper metrics to assess their potentials will be developed.

**Acknowledgment**

The authors thank the NSERC of Canada for their financial support under Grant CRDPJ 424371-11 and also under the Canada Research Chair in Sustainable Smart Eco-Cloud.